\documentclass[epj]{svjour}
\usepackage{graphics}
\begin{document}
\title{Diffractive Vector Meson Production with a Large Momentum Transfer}
\author{Rikard Enberg\inst{1} \and Jeff Forshaw\inst{2} \and 
Leszek Motyka\inst{3} \and Gavin Poludniowski\inst{2}
}                     
\institute{
High Energy Physics, Uppsala University, Box 535, SE-751 21 Uppsala, 
Sweden \and 
Department of Physics \& Astronomy, University of Manchester, 
Manchester M13 9PL, UK \and
Institute of Physics, Jagellonian University, Reymonta 4, 30-059 Krak\'ow, 
Poland
}
\date{Received: date / Revised version: date}
%
\abstract{
We summarise recent progress in the computation of helicity
amplitudes for diffractive vector meson production at large momentum
transfer and their comparison to data collected at the HERA collider.
} 
\maketitle
\begin{figure}
\begin{center}
\resizebox{0.35\textwidth}{!}{
\includegraphics{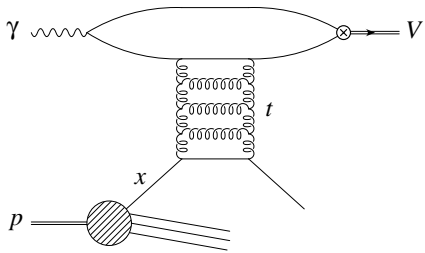} }
\end{center}
\caption{Diffractive vector meson photoproduction at large momentum transfer.}
\label{fig1}
\end{figure}

We are interested in the process illustrated in Figure \ref{fig1} where a
proton and photon collide at high centre-of-mass energy to produce a
vector meson which is far in rapidity from the other final state particles.
So that we can make use of QCD perturbation theory, we insist that the
meson be produced at large transverse momentum. The HERA experiments have
measured the meson $p_T$ spectrum and spin density matrix elements for
the $\rho$, $\phi$ and $J/\psi$ mesons \cite{ZEUS1,Aktas:2003zi}. There 
has also been considerable theoretical interest \cite{FR}-\cite{IK2}.

\begin{figure}
\resizebox{0.5\textwidth}{!}{
\rotatebox{-90}{
\includegraphics{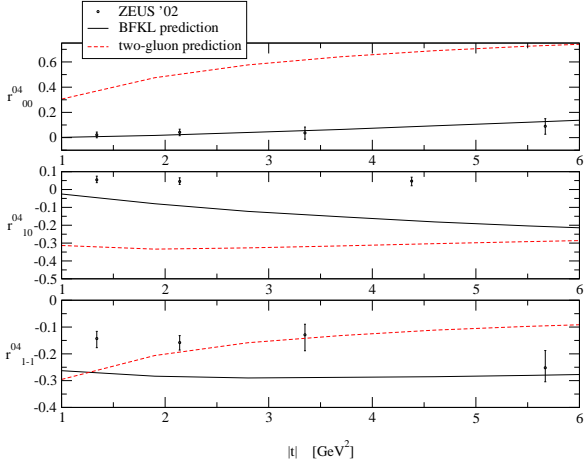} }}
\caption{$\rho$ photoproduction: spin density matrix elements.  
Comparison of the two-gluon exchange prediction
(fixed and running coupling) with BFKL exchange (fixed coupling).}
\label{sdmrho}
\end{figure}

At first sight the data are puzzling. For light quarks one naively expects 
the meson to be predominantly longitudinally polarized and transversely
polarized meson
production to be suppressed by the current quark mass. This is not what
is seen in the data: the meson is without doubt predominantly transversely
polarized for both the $\rho$ and $\phi$ mesons. This can be seen in 
Figure \ref{sdmrho} which shows the measured spin density matrix 
elements for the $\rho$. The results for the $\phi$ 
are very similar to those for
the $\rho$ and we neglect to show them here. 
Writing the helicity amplitudes as $M_{\lambda_{\gamma} \lambda_V}$, 
$r_{00}^{04}$ is proportional to $|M_{+0}|^2$, $r_{10}^{04}$ 
measures the interference between $M_{++}$ and $M_{+0}$, and 
$r_{1-1}^{04}$ measures the interference between $M_{++}$ and $M_{+-}$. 
The challenge is to explain the largeness of $r_{1-1}^{04}$ and the 
smallness of $r_{00}^{04}$.  

Here we report specifically on the results presented in \cite{EFMP1,EFMP2}.
The scattering is supposed to proceed by exchange of a pair of 
interacting reggeized
gluons; corresponding to the sum of all leading logarithms 
$\sim (\alpha_s \ln(s/(-t)))^n$ where 
$-t = p_T^2$ and $s$ is the Mandlestam variable for the hard subprocess. 
Being a leading logarithmic summation, the normalisation of the resulting
amplitudes is not certain, nor is the correct way to treat the strong
coupling $\alpha_s$. Nevertheless, the leading logs do crucially include as
a subset a sum of double logarithms which ensures that the dynamics are
dominated by configurations where the two exchanged gluons share the
momentum transfer. This is to be contrasted with the fixed order 
perturbation theory result which anticipates large contributions from 
asymmetric configurations where one gluon carries all of the momentum
transfer. Summing the leading logarithms also has the virtue that the
amplitude is finite even in the massless quark limit. This is not the 
case for fixed order perturbation theory which is plagued by divergences
which arise from the end-points of the integration over the light-cone
momentum fraction of the quark which forms the meson.

\begin{figure}
\resizebox{0.5\textwidth}{!}{
\rotatebox{-90}{
\includegraphics{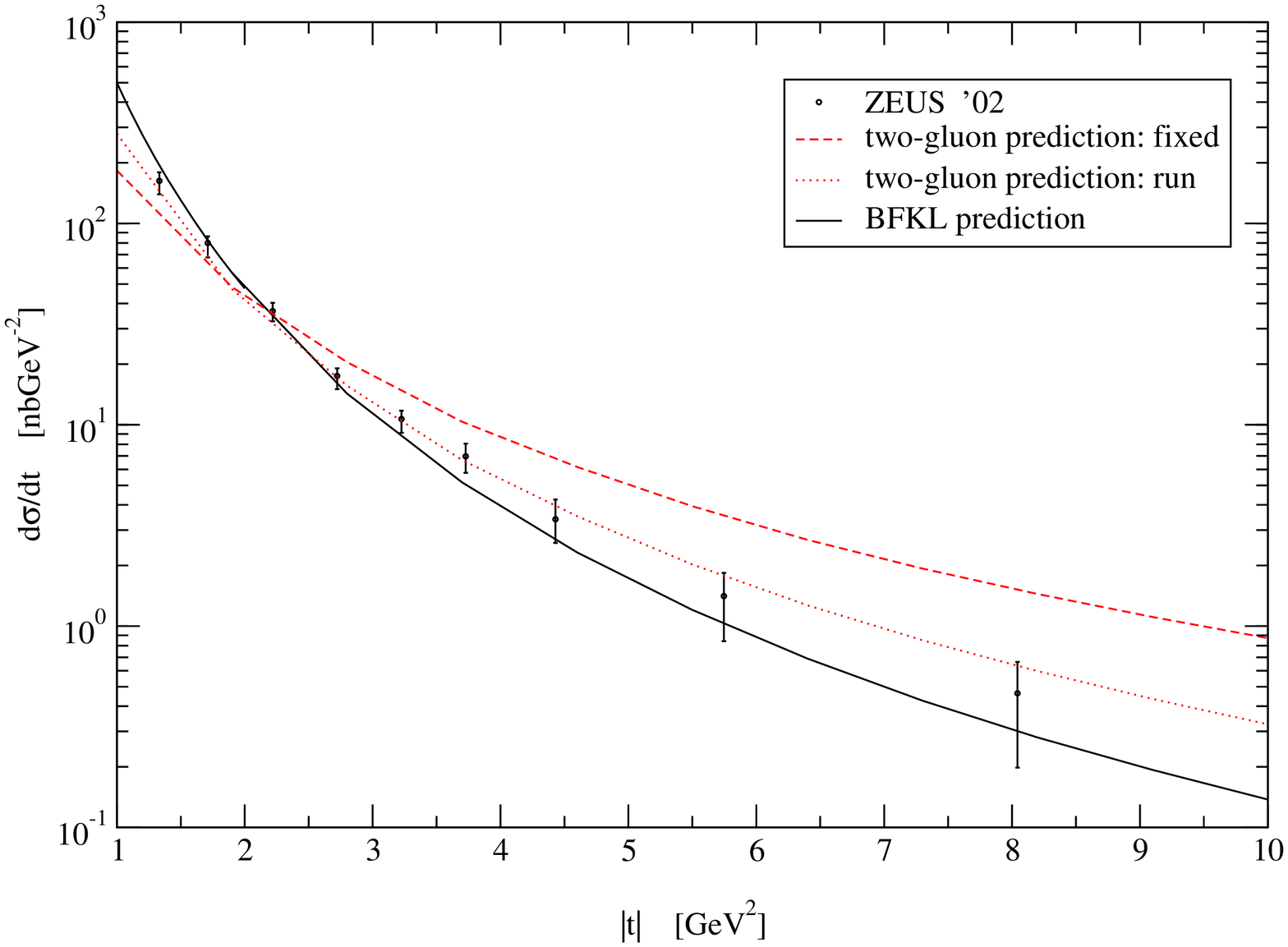} }}
\caption{$\rho$ photoproduction: $t$-distribution.  
Comparison of the two-gluon exchange prediction
(with fixed and running coupling) with BFKL exchange (fixed coupling).}
\label{strho}
\end{figure}

\begin{figure}
\resizebox{0.5\textwidth}{!}{
\rotatebox{-90}{
\includegraphics{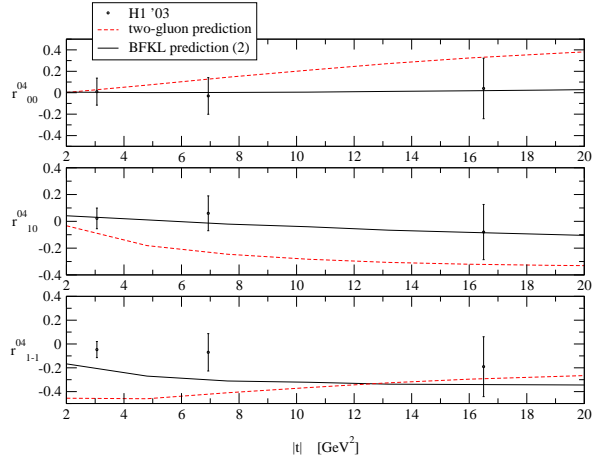} }}
\caption{$J/\psi$ photoproduction: spin density matrix elements.  
Comparison of the two-gluon exchange prediction
(fixed and running coupling) with BFKL exchange (fixed coupling).}
\label{sdmpsi}
\end{figure}

\begin{figure}
\resizebox{0.5\textwidth}{!}{
\rotatebox{-90}{
\includegraphics{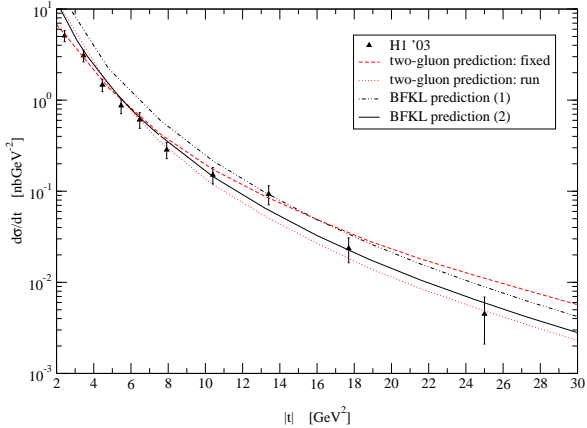} }}
\caption{$J/\psi$ photoproduction: $t$-distribution.  
Comparison of the two-gluon exchange prediction (fixed and running coupling) 
with BFKL exchange (fixed coupling).}
\label{stpsi}
\end{figure}

The production of the meson factorizes from the hard scattering and the
relevant hadronic matrix elements are expanded on the light-cone, as
in \cite{BBKT}. We expand all matrix elements to twist-3, i.e. 
next-to-leading twist. A similar factorization can be perfomed for the photon,
to ensure a clean separation between long and short distances. We do not
however follow this path and instead use the QED coupling of the photon
to the quark and antiquark pair. The quark mass then sets the
size of the chiral odd contributions and also cuts off the integrals
at sufficiently large transverse separations. However, we re-iterate that 
all our amplitudes are finite even in the massless quark limit. In order
to induce a large enough transverse contribution, we use the constituent rather
than current quark mass in the hard scattering amplitude, i.e. we take
$m = m_V/2$ were $m_V$ is the meson mass. The results of the complete
leading logarithmic summation are presented in Figures 
\ref{sdmrho}--\ref{stpsi}, where they are compared to the results obtained
at lowest order in perturbation theory, i.e. corresponding to exchange of
two gluons. 

The two gluon exchange curves are shown for both constant $\alpha_s$
($\alpha_s = 0.27$ for the $\rho$ and $\alpha_s = 0.23$ for the $J/\psi$) 
and running $\alpha_s$ 
($\alpha_s(1~\mathrm{GeV}) = 0.30$ for the $\rho$ and 
$\alpha_s(1~\mathrm{GeV}) = 0.29$ for the $J/\psi$).\footnote{The $\alpha_s$
dependence cancels in the spin density matrix elements.} 
The BFKL (i.e. leading logarithmic) curves are determined using a 
fixed $\alpha_s=0.17$ in the coupling to the external particles and a 
fixed but different $\alpha_s=0.25$ in the BFKL exponent. The same values
are chosen for all mesons. One must also 
choose the scale $\Lambda$ which enters the leading logarithms, 
i.e. $\ln(s/\Lambda^2)$. For all BFKL curves $\Lambda^2 = m_V^2 - t$.
The two sets of BFKL curves labelled $(1)$ and $(2)$ correspond to slightly
different meson tensor decay constants (this choice does not affect the
spin density matrix elements). We refer to \cite{EFMP2} for details on the 
the meson wavefunction and on the effect of making different 
choices for the unknown parameters listed above.
Our main conclusions do not depend upon these details. 

We find that it is generally not too difficult to find a fit for the
meson $p_T$ spectra shown in Figures \ref{strho} and \ref{stpsi}. Both
BFKL and two-gluon exchange can describe the data with rather natural
parameter values. The challenge is to simultaneously describe the data on
the spin density matrix elements. 

For light meson production, the two-gluon exchange
results fail to predict the dominance of transverse meson production even
using the constituent quark mass. The two-gluon exchange predictions are
also plagued by the fact that the quark mass is an essential infrared
regulator. In contrast, BFKL does anticipate predominantly transverse 
meson production and is infra-red finite. However we
did not succeed in finding agreement between BFKL and the $r_{10}^{04}$
matrix element -- BFKL being too large and negative. The double helicity
flip amplitude enters into $r_{1-1}^{04}$ and vanishes at leading twist.
Our next-to-leading twist calculation is therefore only leading order
in this quantity. Using BFKL we tend to overestimate its magnitude. 
For the heavier
$J/\psi$ meson, BFKL is in very good agreement for both the $p_T$ spectrum 
and the spin density matrix elements, although the data do have larger
uncertainties than for the light mesons. 

Looking in detail at the theoretical calculations, one finds that there
are large contributions from the end-point regions of the integral over
the quark (and antiquark) light-cone momentum fractions. It is these 
regions which cause the two-gluon calculation to diverge, and even in the
BFKL case there remain sizeable contributions. Large end-point contributions
bring into question the validity of the factorization of the amplitude
into a perturbative part and a non-perturbative meson matrix element and so
should be a cause of concern. One can rather artificially
suppress these contributions by raising the quark mass and it is noticeable
that after so doing the agreement with the spin density matrix elements
improves substantially. Hitherto, we have ignored the fact that there is
a Sudakov suppression of radiation off the quark and antiquark since we are
dealing with an exclusive quantity. We suggest that this suppression of
radiation may diminish the contribution from the large dipoles which arise
as one moves into the end-point region although it remains to quantify 
the effect.

In summary, the HERA data on diffractive meson production at high $p_T$
are proving a real challenge to explain. Both fixed order and all order
calculations can explain the $p_T$ spectra of the mesons but neither
can at present provide a satisfactory explanation of the helicity structure.
There is room for improvement in the theoretical analyses whilst on the
experimental side the goal must be to obtain data out to larger values of
$t$ for the light mesons, and to reduce the errors, especially for the
$J/\psi$ meson. 

\section*{Acknowledgements}
We thank the organizers of the Aachen EPS conference for providing the
opportunity to present this work. This research was funded in part by the UK 
Particle Physics and Astronomy Research Council (PPARC), by the Swedish 
Research Council, and by the Polish Committee for Scientific Research (KBN) 
grant no.\ 5P03B~14420.


%
%

\end{document}